\documentclass{article}
\usepackage{graphicx,graphics}
\usepackage{url}

\begin{document}

\title{Appearance of an accretion disk perturbed by fractional Brownian Motion density}
\author{G. Mocanu$^{*,+}$ and N. Magyar$^*$ and A. Pardi$^*$ and A. Marcu$^*$}

\date{}

\maketitle

$^*$Faculty of Phycis, Babes-Bolyai University, Cluj Napoca, Romania, No. 1 Kolgalniceanu Street, 400084; $^+$Department of Mathematics, Technical University Cluj Napoca,Memorandumului Street 28,400114 Cluj-Napoca,Romania

\begin{abstract}
This paper aims to investigate/map the effects that perturbations
applied to an accretion disk might produce on the registered Light
Curves (LC). The case of accretion disks around supermassive
active black holes (AGNs) is studied with the goal to explain some of the
statistical properties of the observed IntraDay Variability (IDV). The region producing optical IDV is perturbed by allowing it to develop a mass density of a fractional Brownian Motion-like type. The light curves and
spectral slopes are calculated and compared to observational data for different Hurst parameters.
The spectral slopes of the simulated light curves vary in the range $(0.4,2.5)$. The agreement with observational data shows that a magnetized
disk subjected to stochastic perturbations can produce some of the
features observed in the light curves.
\end{abstract}

\section[]{Introduction}

Extensive observational and theoretical efforts have been made in
order to explain IntraDay Variability (IDV) in some classes of
Active Galactic Nuclei (AGN)~\cite{art:1k,art:11w,art:3p,art:15m,art:18b}. IDV
manifests as the fast change (in less than one day) in the
luminosity output of an object, and for supermassive black holes
IDV occurs in the optical domain. The Power Spectral Distribution (PSD) of these light curves
are found to be nontrivial~\cite{art:25c,art:9a,art:12l}. A large percentage
of AGN evolve so as to produce IDV in the optical with PSD for
which the spectral slope $\alpha$ is neither $0$ nor $-2$, as
shown by structure function analysis~\cite{art:25c}, fractal dimension
analysis \cite{art:12l} and discrete Fourier transform analysis~\cite{art:9a}. 

These types of analyses lead to the conclusion that the disk is perturbed by a stochastic noise; further discussion on this topic can be found in, e.g., \cite{art:50m} which find no convincing case of periodicity in LCs for a 9000 quasar sample and state that these type of oscillations are entirely expected from coloured noise processes; \cite{art:29m} conclude, in a discussion of optical IDV in Bl Lacs that IDV is essentially a stochastic process; \cite{art:12d} state that it is unlikely that the observed quasar variability is caused by a coherently varying accretion disk.

For BL Lac S5 0716+714 the spectral slope was found to vary from $1.083$ to $2.65$ for the IDV data set discussed in~\cite{art:PSD} and~\cite{art:3p}, from $0.6$ to $2.22$ for the data set discussed in~\cite{art:25c}, from $0.8$ to $1.4$ in the data set discussed in~\cite{art:9a} and from $0.6$ to $1.7$ in the data set discussed in~\cite{art:12l}.

We start from the assumption that the source of IDV is placed
within a geometrically thin, optically thick disk. An equilibrium rotating
accretion disk in the MHD framework is perturbed (Section~\ref{sect:Perturbation}) by a fractional Brownian Motion (fBM) process in density. The MHD equations
are solved to obtain the perturbed variables needed to compute the
Reynolds-Maxwell stress tensor component responsible for the
emitted luminosity. The luminosity is calculated by integrating this stress component over a radius between $[5,20]$ gravitational radii.

Although in a non-complicated approach, this work aims to advance the study in this area by using a fractional process as a perturbation, superimposed on a magnetized medium. As it is shown in the body of the paper (Section~\ref{sect:config2}), an equilibrium deterministic magnetic field acquires a stochastic component following such a perturbation. 

In the Conclusions (Section~\ref{sect:Conclusions}) we review the
signature of the input fBM on the output luminosity and its
connections to observed light curves.

\section[]{Disk configuration and equilibrium\label{sect:Equilibrium}}

The initial configuration consists of a geometrically thin, optically thick rotating magnetized disk. By assumption the disk is non-viscous. The fundamental equations for accretion disk structure (e.g. Equation set 6 from~\cite{art:18b}) are the continuity equation

\begin{equation}
\frac{\partial \rho}{\partial t} + \nabla \cdot (\rho \vec{v}) = 0,
\end{equation}
the equation of motion for a magnetized plasma element

\begin{equation}
\rho \frac{\partial \vec{v}}{\partial t} + \left ( \rho \vec{v} \cdot \nabla \right ) \vec{v} = -\nabla P - \rho \nabla \phi _g  + \frac{1}{\mu _0} \left ( \nabla \times \vec{B} \right ) \times \vec{B},
\end{equation}
the induction equation

\begin{equation}
\frac{\partial \vec{B}}{\partial t} = \nabla \times \left ( \vec{v} \times \vec{B} \right )
\end{equation}
and the zero-divergence condition imposed on the magnetic field

\begin{equation}
\nabla \cdot \vec{B} = 0,
\end{equation}
where $\rho$ is the volume density, $\vec{v}$ is the velocity of the plasma element, $P$ is the hydrostatic pressure in the disk, $\phi _g$ is the gravitational potential due to the central source and $\vec{B}$ is the magnetic field permeating the disk.

Also, we use the assumption that the plasma in the disk is an ideal gas in the sense that the equation

\begin{equation}
p = \frac{k_B}{\tilde{\mu} m_p} \rho T
\end{equation}
is valid, where $k_B$ is the Boltzmann constant, $\tilde{\mu}$ is the mean molecular weight, $m_p$ is the proton mass and $T$ is the temperature.

The equilibrium disk is well described by a time steady
cylindrical symmetry $\left (r, \phi, z \right )$, such that $\partial /\partial t = \partial /
\partial \phi = 0$. The disk structure may be characterized by a
central temperature depending only on $r$ and the dependence on
the $z$ coordinate may be neglected~\cite{art:12s}. As a feature of
the standard thin disk assumption, all the quantities should be considered as averages over height.

The gravitational potential outside the event horizon of the super
massive black hole is described by the Newtonian potential
$\phi_{g} (r)= -\frac{GM}{r}$, where $G$ is the gravitational
constant and $M$ is the mass of the black hole. A test plasma
volume in a Keplerian disk experiences a velocity equal to the
Kepler velocity  $\mathbf{v}_0 (r)= \left ( 0, u_k (r), 0 \right
); u_k (r) = \sqrt{GM/r}$, and an angular velocity equal to the
Kepler angular velocity $\mathbf{\Omega}_0 (r) = \left ( 0, 0,
\Omega_k (r) \right ); \Omega_k (r) = \sqrt{GM/r^3}$.

The equilibrium magnetic field is set up to have a zero radial component. Observations place the
value of the magnetic field in the disk around the modest value of $\approx 10^{-8\pm1}$T~\cite{art:21k}.

The numerical values chosen for the boundary parameters do no
necessarily satisfy the zero-torque boundary condition on the inner orbit.
Historically, the stress on the most inner orbit of the
disk was considered to be zero but recent discussions on this
topic have led to the conclusion that this assumption must be
relaxed. For example, in recent work, \cite{art:20a} numerically
investigate a disk by taking into account general relativistic
effects, non-zero torque at the boundary  and find that the
radiation returning from the inner area (which was previously
decoupled from the disk because of the no-torque boundary
condition) causes various annuli in the disk to "communicate" on
the light crossing timescale. Communication on this timescale is a
feature needed by any model trying to explain IDV, because
observational data show that if the source of IDV is in the disk,
it needs to propagate on these timescales~\cite{art:rmsflux,art:12d}.

Similar to the standard disk model~\cite{art:12s,art:26s}, the height averaged equilibrium density as $\rho _0 (r) \sim r^{-\theta}$ with $\theta = 15/8$ and the equilibrium
temperature is taken as $T_0(r) = k r^{-\tau}$. This is
justified based on the self similarity property of the disk
solutions~\cite{art:10b} and on a set of observations. For example,
the optical/UV continuum of NGC 7469 corresponds with the
prediction for a $T_{eff} \sim r^{-3/4}$~\cite{art:32k}. \cite{art:12g} reports on
a set of observations where the observed optical-UV spectral
energy distribution (SED) implies a temperature $T \sim
r^{-0.57}$, independent of the thickness of the disk. Simple black
body disk models around super-massive black holes have $T_{max}$
less than $3-8\times 10^5$K and the UV and optical continuum
emitting region is placed in the region between
$10-100r_g$~\cite{art:32k,art:18b}, where $r_g$ is the gravitational radius $r_g=2GM/c^2$. For a central mass
of $M=10^8 M_\odot$, the accretion flow region responsible for the
observed optical/UV spectrum is $\approx 10r_g$~\cite{art:32k} and
$T\approx 10^5K$ is a characteristic temperature within about
$30r_g$~\cite{book:10f}.

Based on these arguments, the numerical values of the parameters
used throughout this analysis (all expressed in SI units) are

\begin{itemize}

\item{}physical constants

\begin{eqnarray}
&&\widetilde{\mu}=1.27, \mbox{   }c=3\times 10^8, \mbox{
}M_{\odot}= 2 \times 10^{30}, \mbox{
}G=6.6\times10^{-11},\nonumber \\&&k_B=1.3\times10^{-23},\mbox{
}\mu _0 = 4\pi10^{-7},\mbox{ }m_p=1.6\times10^{-27}
\end{eqnarray}

\item{}model assumptions

\begin{eqnarray}
&&\tau = -3/4,\mbox{ } M=10^8M_\odot,\mbox{ }r_{o}=20r_g,\mbox{ }r_{i}=5r_g,\nonumber \\&&\theta = 15/8,\mbox{ } k=2.65 10^{14}.
\end{eqnarray}

\end{itemize}
where $r_o$ and $r_i$ are the outer and inner radii of the section of the disk considered to produce the optical radiation. 

\section{Density Perturbation}\label{sect:Perturbation}

The equilibrium state is perturbed by a prescribed temporal variation in the
density. When the perturbation is applied,
the new physical parameters that satisfy the MHD equations
consist of a sum of two parts: the equilibrium part and the perturbed part.

The perturbations are isothermal. This is not
because we believe this is necessarily the case in an
astrophysical disk. We make this assumption because in this toy
model we want to study fluctuations in energy occurring at the
same temperature, and not the distribution of energy in a large
interval of temperatures.

Accretion is possible due to the outward angular momentum transport by a sum of Reynolds and
Maxwell stresses~\cite{art:21b,art:11b,art:18b} and we assume that the
perturbations in this mechanism are responsible for the
variability. The component of the stress tensor responsible for
the luminosity is defined as

\begin{equation}
 m_{r\phi} (r,t) = \mu _0 \rho _0 (r) v_{r} (r,t) v_{\phi} (r,t) - B_{r} (r,t) B_{\phi} (r,t),
\end{equation}
where $v_r, v_\phi, b_r, b_\phi$ are the perturbed components of the velocity and magnetic field.

The luminosity emitted from a patch of the disk due to the
perturbation, between radii $r_1$ and $r_2$ is~\cite{art:21b}

\begin{equation}
L_{12} (t) = \int _{r_1} ^{r_2} m_{r\phi} (r,t) dr
\end{equation}
and this equation is the starting point of producing the light curves shown in the Results section of the paper.

\subsection{Stochastic density perturbation\label{sect:densFBM}}

There is a general consensus that the underlying mechanism
producing IDV is stochastic~\cite{art:29l,art:16a,art:50m}. 

\begin{figure}
\begin{center}
\includegraphics*[width=6cm,angle=0]{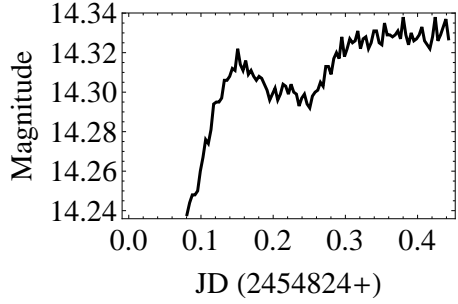}
\end{center}
\caption{B (blue) band time series for the object Bl Lacertae S5 0716+714; JD stands for Julian Day.\label{fig:idv}}
\end{figure}

The time dependency of the density perturbation is taken to be the realization of a stochastic process such that $\frac{\partial \rho _1}{\partial t} = A_\rho B_H (t) = \xi$, where $B_H (t)$ is a normalized fractional Brownian Motion (fBM) and $A_\rho$ is a constant amplitude, which will be prescribed by us. It is expected that locally
fluctuating component of magnetic field create numerous current
sheets inside the disk~\cite{art:17m}, which can be used as random
reconnection sites. The accretion disk can work as a dynamo
amplifying the stochastic magnetic field~\cite{book:22b}. This happens
because even in a perfectly conducting plasma, turbulence develops
through the MagnetoRotational Instability (MRI)~\cite{art:42s}. The already mentioned peculiar shape of the observational PSD
prompts us to consider that the perturbation is a signal with
non-Markovian properties. A sample light curve is shown in Fig.~\ref{fig:idv}.

There is no constraint imposed to the value of $H$ in our framework mainly because there is a lack of observational data to correlate IDV characteristics to characteristics of the central engine. For example, \cite{art:7n} report, regarding the optical IDV  of S5 0716+714, that there is no correlation between the source magnitude and the amplitude of the IDV and that there is no clear correlation between the source magnitude and the rate of magnitude variation; \cite{art:5w} report that time scales of optical variability in BL Lacs do not correlate with luminosity; 
\cite{art:50m} conclude that it cannot be assumed that quasars with similar mass will necessarily have the same variability properties; \cite{art:7h} discuss optical data of a sample of Bl Lac and state that there are no dependencies of the time scales on intrinsic properties, redshift or absolute magnitude.

The \emph{definition} of fBM is~\cite{art:52m}

\begin{eqnarray}
\lefteqn{B_H(t) = \frac{1}{\Gamma \left ( H +
\frac{1}{2}\right)}\left( \int _{-\infty}^0 \left [
(t-s)^{H-1/2}-(-s)^{H-1/2}\right ] dB(s) \right)}\nonumber\\
&& +\frac{1}{\Gamma \left ( H + \frac{1}{2}\right)}\left(\int _0^t
(t-s)^{H-1/2}dB(s)\right),
\end{eqnarray}
where $\Gamma$ represents the special function

\begin{equation}
\Gamma(z) = \int _{0}^{\infty} x^{z -1}e^{-x}dx. \\
\end{equation}

The Power Spectral Distribution of such a signal depends on the Hurst parameter $H$ as~\cite{art:52m}

\begin{equation}
P(f)\sim f^{1-2H}.\label{eq:psdH}
\end{equation}

If time is discretised as $t=hk$, with $h$ a constant timestep, the sequence of increments

\begin{equation}
G_H (k) = B_H (k) - B_H(k-1)
\end{equation}
is what is sometimes called fractional Gaussian noise, with the
property that it is causally strongly connected with itself even
at infinite temporal range. This is seen by calculating the
expectation value of the correlation function of this signal in
the limit of an infinite numbers of timesteps

\begin{equation}
\lim _{k\to \infty} E\left [ G_H(j)G_h(j+k)\right] \sim
H(2H-1)k^{2H-2} \neq 0 
\end{equation}
for $H \notin \{0,1/2\}.$ So stated more clearly, the number $H$ is an indication of the memory of the process: the fBM $B_H$ has increments that are not independent of each other.

The set $\{ v_r, v_\phi,
B_r, B_\phi \}$ at each position and timestep is needed. The
necessary MHD equations are written as stochastic differential
equations, with $\xi (t)$ as a source term. 

\subsection{Equations valid for general $B_0(r)=\left \{0,B_\phi(r),B_z(r) \right\}$}

After linearising the perturbed set of MHD equations and selecting just the needed set of variables, it is obtained that

\begin{equation}
\frac{\partial \rho _1}{\partial t} + \frac{\rho _0}{r}\frac{\partial (rv_r)}{\partial r} + v_r\frac{d\rho_0}{dr} = 0,
\end{equation}

\begin{equation}
\frac{\partial v_\phi}{\partial t} + v_r \left (\frac{u_k}{r}+\frac{du_k}{dr}\right) = \frac{1}{\mu _0 \rho _0}\frac{a_1}{r^2}\frac{d(rB_\phi)}{dr},
\end{equation}

\begin{equation}
\frac{\partial b_\phi}{\partial t} = a_1 \frac{d(u_k/r)}{dr}-\frac{\partial (v_r B_\phi)}{\partial r},
\end{equation}

\begin{equation}
b_r = \frac{a_1}{r}.
\end{equation}

The continuity equation provides $v_r$ in terms of $\xi (t)$

\begin{equation}
\rho _0 (r) \frac{\partial v_r (r,t)}{\partial r} + \left [
\frac{\rho _0 (r)}{r} + \frac{d \rho _0 (r)}{dr} \right ] = - \xi
(t),
\end{equation}
which when solved for the $r$ variation with $v_r(r_o,t)=0$ gives

\begin{equation}\label{eqS:vr}
v_r (r,t) = \frac{r^{7/8} (-r^2 + r_o^2) }{2 a_2} \xi (t).
\end{equation}

\subsection{Configuration I of initial magnetic field\label{sect:config1}}

We study how a disk with a particular initial configuration will evolve when subjected to a stochastic density perturbation. We let $B_\phi = 0$ and

\begin{equation}
B_z(r) = \frac{b_0}{r}\sin \left [ \frac{2\pi (r-r_i)}{r_o-r_i} \right ]
\end{equation}

The value of $b_0$ is calculated by fixing the plasma beta $\beta = 2 p_0 \mu_0 / B_0^2$ to a numerical value.

The $\phi$ component for the equation of motion and induction simplify and become

\begin{equation}
\frac{\partial v_\phi}{\partial t} + v_r \left (\frac{u_k}{r}+\frac{du_k}{dr}\right) = 0,
\end{equation}
and

\begin{equation}
\frac{\partial b_\phi}{\partial t} = a_1 \frac{d(u_k/r)}{dr}.
\end{equation}

The solution for the magnetic field is, with $b_\phi(r,t=0)=0$

\begin{equation}
b_\phi (r,t) = -\frac{3a_1}{2}\frac{u_k}{r^2} t = - \frac{3a_1}{2}\Omega _0 t.
\end{equation}

For this case we find that the perturbed magnetic field is not a stochastic variable.

The equation for the azimuthal velocity becomes

\begin{equation}
\frac{\partial v_\phi}{\partial t} + \frac{1}{4a_2}u_k r^{-1/8} (-r^2+r_o^2)A_\rho B_H (t)=0
\end{equation}

The stochastic quantities in this configuration are $v_r$ and $v_\phi$ and have defining equations

\begin{equation}
v_r (r,t) = f_1 (r) B_H (t),
\end{equation}

\begin{equation}
\frac{\partial v_\phi (r,t)}{\partial t} = f_2(r) B_H(t),
\end{equation}
with

\begin{equation}
f_1(r) = A_\rho \frac{r^{7/8} (-r^2 + r_o^2) }{2 a_2} 
\end{equation}
and

\begin{equation}
f_2(r) = -\frac{A_\rho}{4a_2}u_k r^{-1/8} (-r^2+r_o^2).
\end{equation}

We take advantage that there is no differential operator in $r$ and write the defining equations in update form for the temporal variation

\begin{equation}
v_\phi (r,t+h) = v_\phi (r,t) + f_2(r) \sqrt{h} G_H (t)
\end{equation}
with $v_\phi (r,0)=0$.

The way that $a_1$ is assigned a numerical value in this case takes advantage of the fact that $b_\phi(r,t=0)=0$, and that $b_r=a_1/r$ such that we start the simulation by imposing that at $t=0$, the perturbed quantities have the same plasma beta as the equilibrium.

\subsection{Configuration II of initial magnetic field\label{sect:config2}}

The shape of $B_z(r)$ is kept and $B_\phi (r) = \frac{b_0}{r}\cos \left [ \frac{2\pi (r-r_i)}{r_o-r_i} \right ]$. $b_0$ is again calculated based on plasma beta considerations.

In this case the evolution equations for $v_\phi$ and $b_\phi$ are

\begin{equation}
v_\phi (r, t+h) = v_\phi (r,t) - f_3(r) h +f_4 (r)\sqrt{h}G_H(t),
\end{equation}

\begin{equation}
b_\phi (r,t+h) = b_\phi (r,t) + f_5 (r) h - f_6 (r)\sqrt{h}G_H(t)\label{eq:bphi2},
\end{equation}
where

\begin{equation}
f_3(r)=-\frac{1}{\mu_0 \rho_0(r)}\frac{a_1}{r^2}\frac{d(rB_\phi(r))}{dr},
\end{equation}

\begin{equation}
f_4(r)= - \left ( \frac{u_k}{r}+\frac{du_k}{dr}\right ) f_1(r),
\end{equation}

\begin{equation}
f_5(r) = a_1 \frac{du_k/r}{dr},
\end{equation}

\begin{equation}
f_6(r) = \frac{dB_\phi f_1}{dr}.
\end{equation}

With the assumptions of the model, it is clear from Eq.~\ref{eq:bphi2} that if there is a non-zero initial azimuthal magnetic field, the perturbed magnetic field will be a stochastic process.

\section{Results}

The equations are implemented and solved in Mathematica~\cite{math}. The numerical values for the luminosity are not expected to fit observed ones. The graphical representation is done by dividing the values in the luminosity vector to its minimum value: one can than see by how much the luminosity increases in a time interval of 30 minutes. Variations on this timescale have been reported for this object by, e.g., \cite{art:24w}.

The simulation was carried for both initial magnetic field configurations (Sections~\ref{sect:config1} and \ref{sect:config2}). The set of variable parameters is $\{\beta, H \}$, the plasma beta characterising the system and the Hurst parameter, characterising the perturbation. We chose to let $\beta$ take values $\in \{ 100,1000\}$ and $H\in [0.1, 0.9]$ in increments of $0.1$. If the disk would not modify the properties of the input perturbation it would be expected that the spectral slope would be the one given by Eq.~\ref{eq:psdH}, i.e., within an interval $(-1,1)$ for $H\in (0,1)$.

Conversely, if we assume that the spectral slope of the observed LCs is in a bijective relation with the Hurst parameter of the input perturbation, than this perturbation would have $H_{0} \in (0.8, 1.6)$.

\begin{figure}
\begin{center}
\includegraphics*[width=8cm,angle=0]{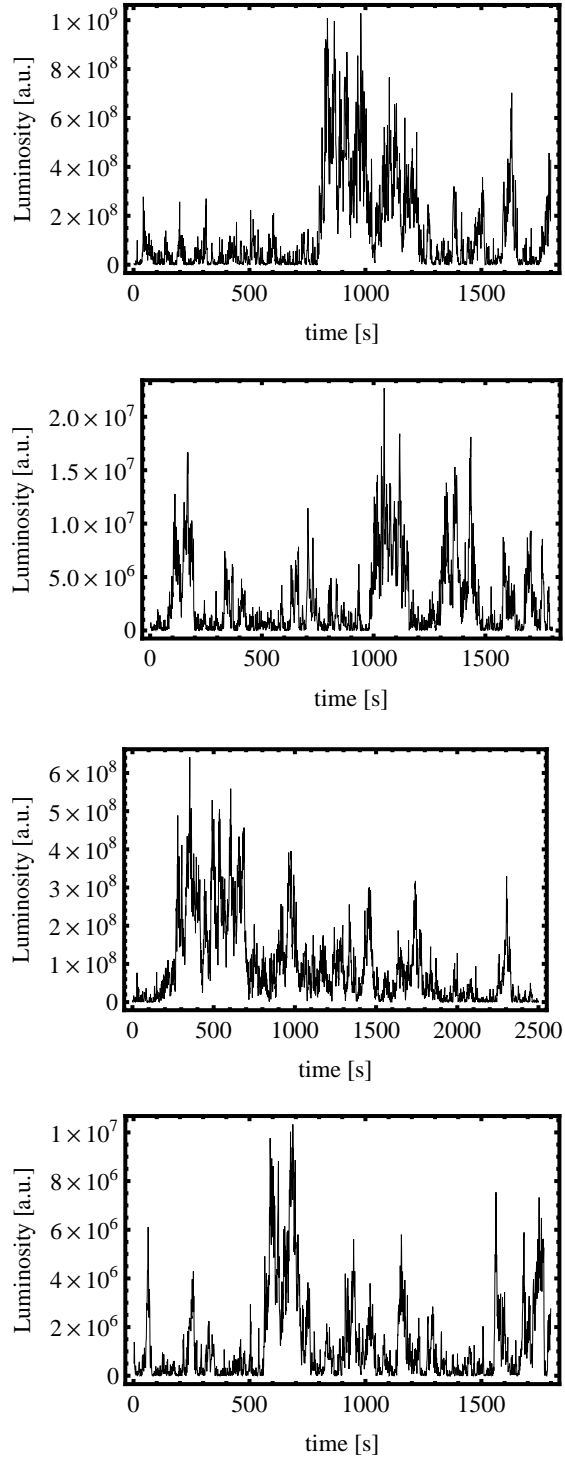}
\end{center}
\caption{Simulated light curves for $H=0.3$, $A_\rho = 10^{-3}$ and, from top to bottom, configuration I $\beta =100$ and $\beta =1000$, configuration II $\beta =100$  and II $\beta =1000$.\label{fig:LCsI}}
\end{figure}

\begin{figure}
\begin{center}
\includegraphics*[width=8cm,angle=0]{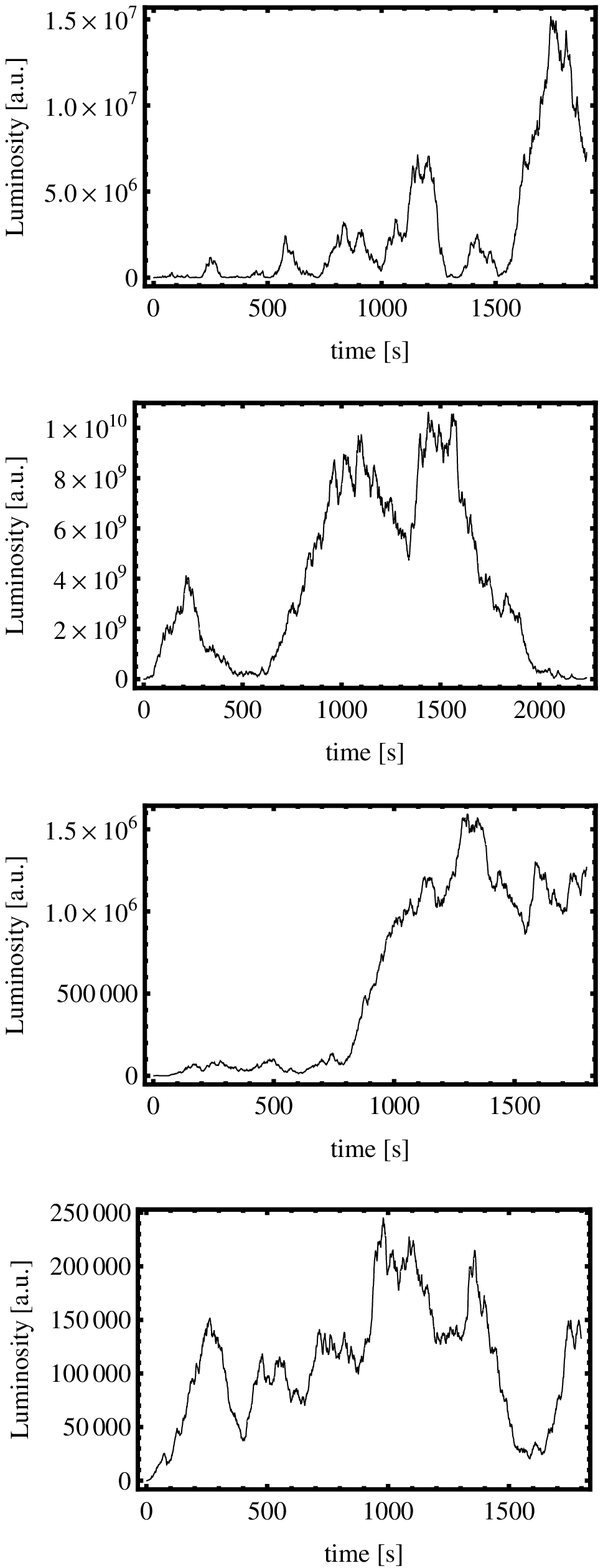}
\end{center}
\caption{Simulated light curves for $H=0.7$, $A_\rho = 10^{-3}$ and, from top to bottom, configuration I $\beta =100$ and $\beta =1000$, configuration II $\beta =100$  and $\beta =1000$.\label{fig:LCsII}}
\end{figure}

\begin{figure}
\begin{center}
\includegraphics*[width=8cm,angle=0]{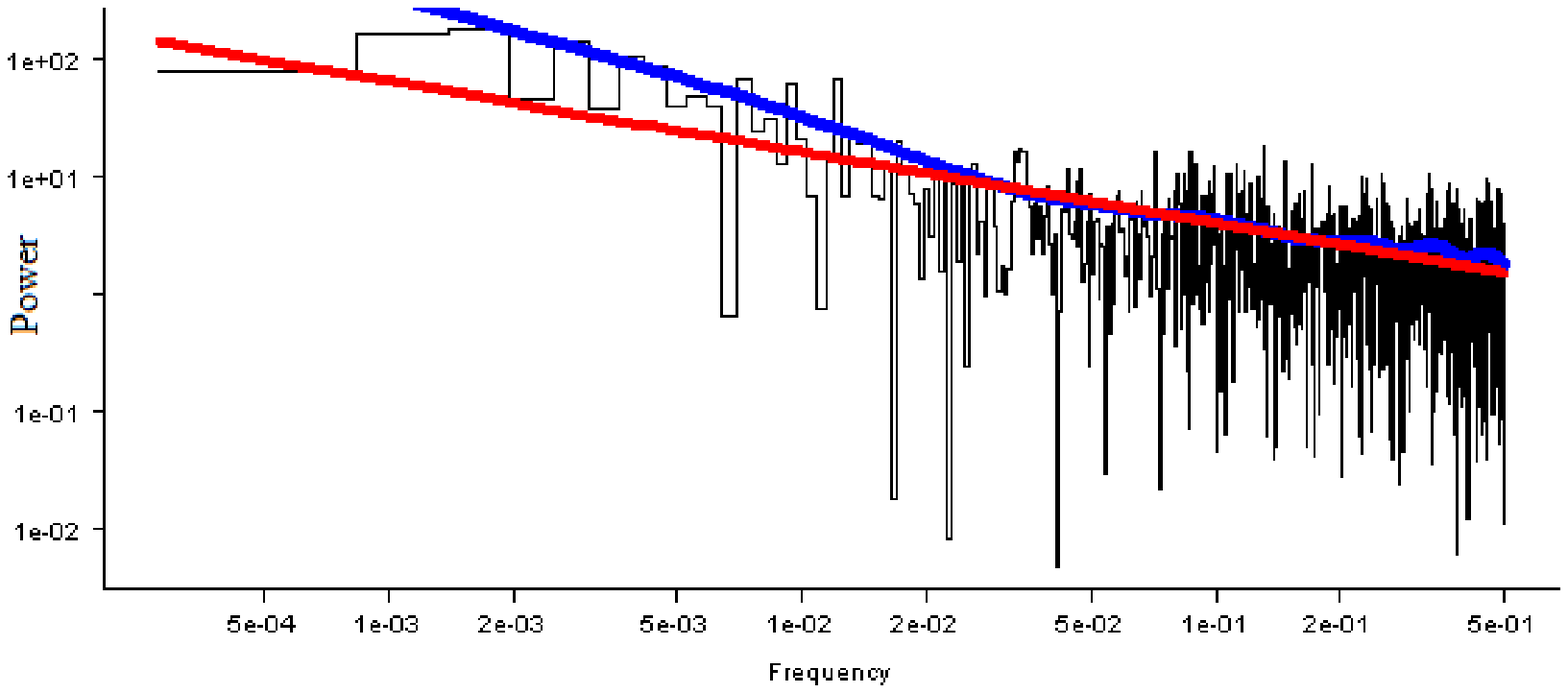}
\end{center}
\caption{Sample PSD output of bayes.R for configuration I, $H=0.1$, $A_\rho = 10^{-3}$ and $\beta =100$.\label{fig:samplePSD}}
\end{figure}

The main results of this work are the light curves produced by varying $\{\beta, H \}$ (a sample of such light curves is shown here in Figs.~\ref{fig:LCsI}-\ref{fig:LCsII}) and most importantly the PSDs of these light curves, all displayed in Tables~\ref{tab:C1b100-5-20}-\ref{tab:C2b1000-5-100}. A sample fit of the PSD is shown in Fig.~\ref{fig:samplePSD}, where the blue line is the best fit and the red line is the fit according to the polynomial assumption. Analysis of IDV observational data for this object place the value of the spectral slope $\alpha$ in a wide interval, ranging from $0.37$ to $2.7$ (see, e.g., Table 2 from~\cite{art:PSD}).

All data vectors were checked for convergence (even though the trend in some figures seems to be increasing), but for consistency of the presentation only the first 30 minutes of all light curves are shown (one timesteps equals one second). Observations show that optical light curves from BL Lac S5 0716+714 although variable are not periodic, e.g. \cite{art:5w} reports that optical observations never have constant duty-cycle; also, \cite{art:24c} found no periodicity during an observational campaign in March 2009.

The tables were populated by using the .R software~\cite{art:3v} according to the assumption that the emission process behaves so as to produce a luminosity with PSD $\sim f^{-\alpha}$. We impose this restriction because, as already discussed, many sets of observational data exhibit this behaviour. The second column in each table contains the value of the spectral slope $\alpha$ and the first column contains the Hurst parameter of the fBM perturbing the disk in that case. The errors associated to each $\alpha$ stem from the fact that the slope is calculated with the .R software; essentially, these errors come from mediating over many realizations of a process with the same characteristics. The third column contains the Bayesian probability which assesses the correctness of the assumption, i.e. if $p_B$ is close to $1$, than the assumption is correct. For details about the technical procedure, see~\cite{art:3v}.

 We would consider our model to be at least moderately successful if it would reproduce non-trivial values for $\alpha$ with high values of Bayesian probability. A careful analysis of Tables~\ref{tab:C1b100-5-20}-\ref{tab:C2b1000-5-100} shows that this is indeed the case; a bold type letter was used to mark the most promising results.

There is one subtle issue that still needs to be discussed, namely what happens to the radiation after the local emission. The mathematics used so far assumed, according to the standard model, that the produced radiation comes from a height averaged quantity; lets call this the "observed" radiation. However, from its production spot in the disk and before leaving the body of the disk, the local radiation goes through diffusion in the thickness (albeit small) of the disk. So the local radiation, upon diffusion, becomes the observed radiation. There would appear to be an inconsistency in our treatment: we use tools to obtain the observed radiation (height averaged equations) but use local perturbation (the fBM). This apparent inconsistency can be dealt with by providing an answer to the question: are statistical properties of the distribution of isothermal photons (which are responsible for the local radiation) changed if they are subjected to unbiased random walk (diffusion through the body of the disk)? Recall that our aim is to offer a possible explanation of the PSD of the observed light curves. Since unbiased random walk through a medium cannot influence temporal correlation of otherwise identical (same temperature) photons, we will use the term "observed" radiation for our results.

Comparison with the statistical properties of observational data for a BL Lac, e.g. analysed in \cite{art:PSD}, is good.

\begin{figure}
\begin{center}
\includegraphics*[width=8cm,angle=0]{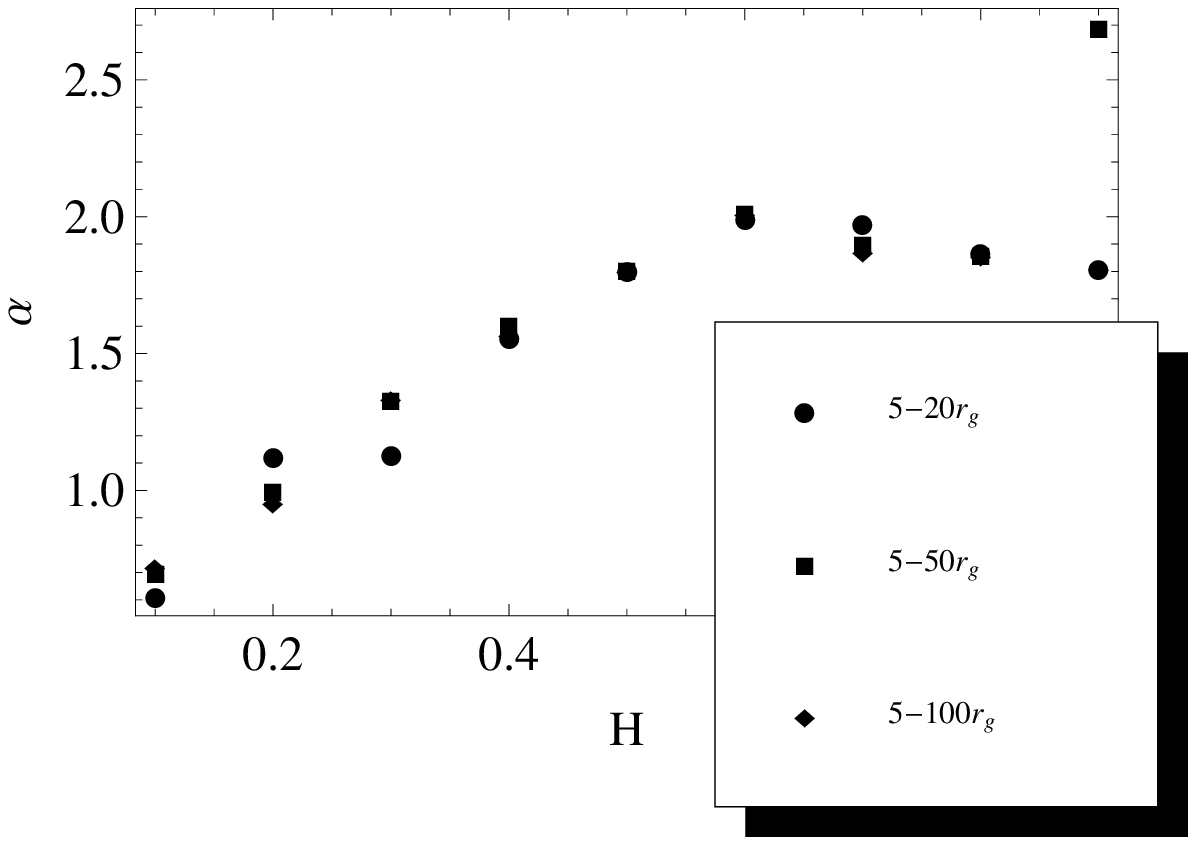}
\end{center}
\caption{A plot of the output spectral slope $\alpha$ as a function of the input Hurst parameter $H\in \{0.1,0.9\}$ for configuration I, $\beta = 100$ and various radial extents of the disk.\label{fig:Ha-1}}
\end{figure}

\begin{figure}
\begin{center}
\includegraphics*[width=8cm,angle=0]{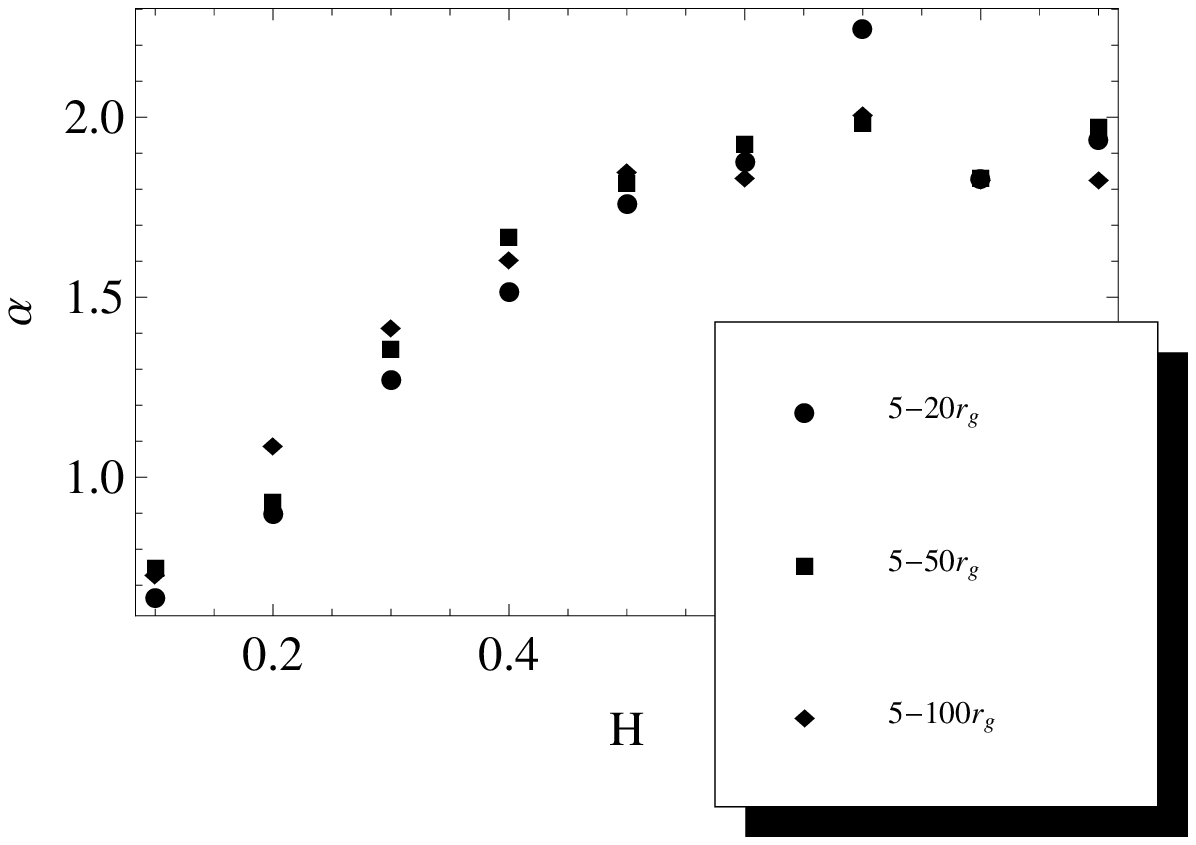}
\end{center}
\caption{A plot of the output spectral slope $\alpha$ as a function of the input Hurst parameter $H\in \{0.1,0.9\}$ for configuration I, $\beta = 1000$ and various radial extents of the disk.\label{fig:Ha-2}}
\end{figure}

\begin{figure}
\begin{center}
\includegraphics*[width=8cm,angle=0]{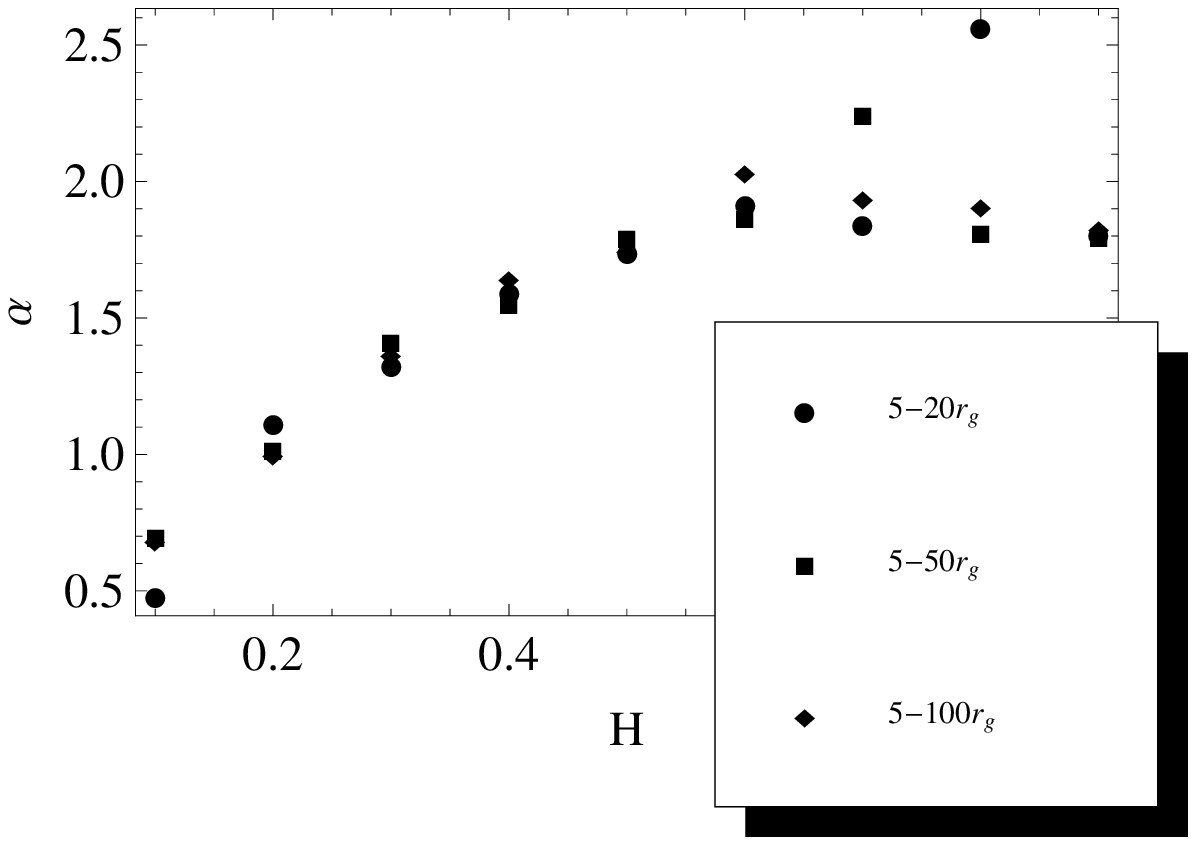}
\end{center}
\caption{A plot of the output spectral slope $\alpha$ as a function of the input Hurst parameter $H\in \{0.1,0.9\}$ for configuration II, $\beta = 100$ and various radial extents of the disk.\label{fig:Ha-3}}
\end{figure}

\begin{figure}
\begin{center}
\includegraphics*[width=8cm,angle=0]{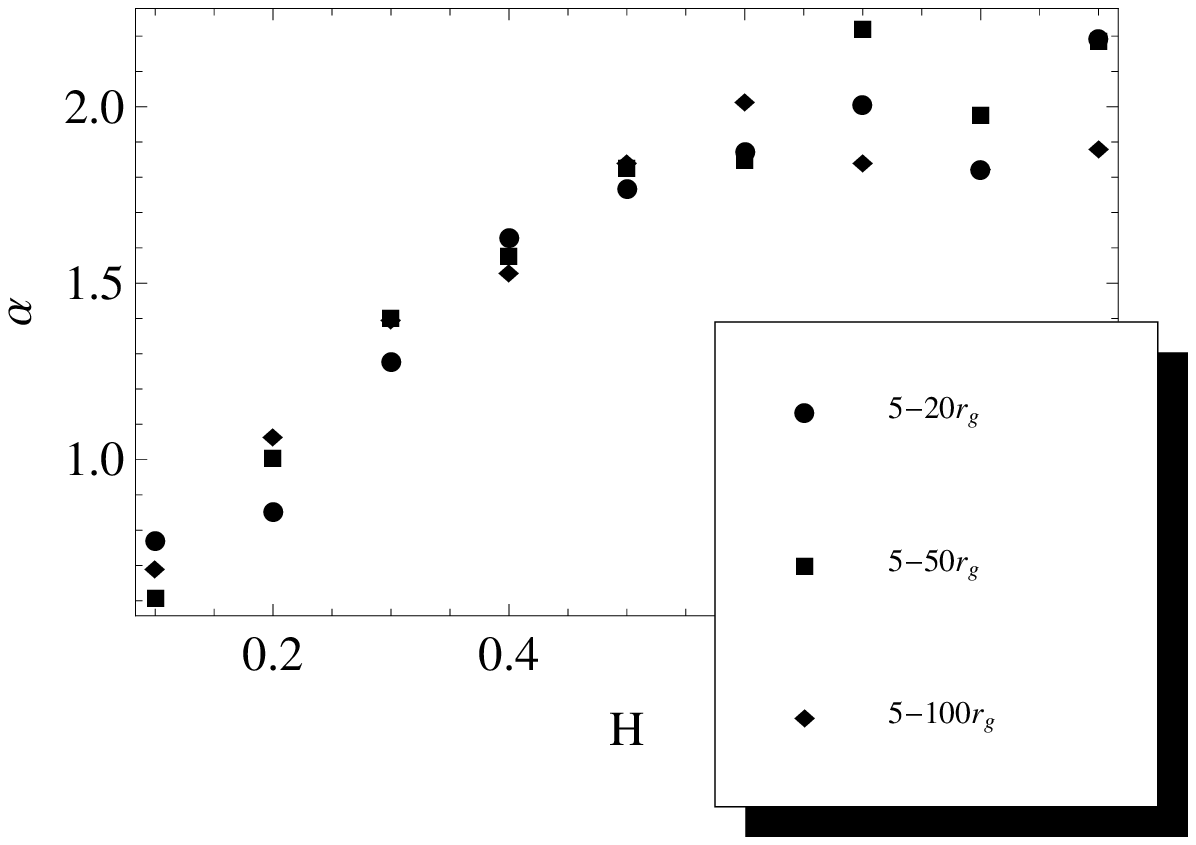}
\end{center}
\caption{A plot of the output spectral slope $\alpha$ as a function of the input Hurst parameter $H\in \{0.1,0.9\}$ for configuration II, $\beta = 1000$ and various radial extents of the disk.\label{fig:Ha-4}}
\end{figure}

Regarding the connection between $\alpha$ and $H$ in the space parameter of the two different configurations, the plasma $\beta$ and different radial extents of the simulated disk, we produced plots $\alpha (H)$ with varying radial extent (Figs.~\ref{fig:Ha-1} - \ref{fig:Ha-4}). It does not seem that extending the disk produces significant changes in the shape of the relationship. As expected, the dependency is not linear. In fact, it can be seen from the plots that $\alpha (H)$ reaches a plateau, with a different value for each of the configurations considered. However, these results should be taken with caution as the points in the plots do not all have equal Bayesian probability (the $p_B$ from the Tables is different).

\section{Conclusions}\label{sect:Conclusions}

The appearance of a perturbed magnetized accretion disk was
investigated for the realistic case of a stochastic
density perturbation, the output was investigated as a function of
the Hurst parameter of the input signal. Two possible configurations of an initial magnetic field were considered. The most important result is the set of values obtained for the spectral slope and the corresponding Bayesian probabilities ($\alpha$ and $p_B$, shown in Tables~\ref{tab:C1b100-5-20}-\ref{tab:C2b1000-5-100}). The simulated curves are found to reproduce the PSD characteristics of observed IDV data.

Although left with important things to fine-tune, this toy model
may turn out to be very useful in explaining IDV. Future research in this area would benefit from focusing on finding a physics-based analytical bijective relation between input perturbation and produced light curves. 

\section*{Acknowledgements}

The authors thank the referee for valuable comments and suggestions which increased the quality and overall accuracy of this paper.

\newpage

\begin{table}
\begin{center}
\caption{PSD of the output curve for a range of Hurst parameters
of the input signal, for Configuration I, $\beta$ = 100, $r\in [5r_g,20r_g]$ \label{tab:C1b100-5-20}}
\begin{tabular}{lll}
\hline
H   & $\alpha$            & $p_B$ \\ \hline
\textbf{0.1} & 0.6111 [$\pm 0.03$] & 0.888  \\ \hline
\textbf{0.2} & 1.1243 [$\pm 0.03$] & 0.688  \\ \hline
0.3 & 1.1298 [$\pm 0.03$] & 0.432 \\ \hline
0.4 & 1.5573 [$\pm 0.03$] & 0.604  \\ \hline
0.5 & 1.8014 [$\pm 0.032$] & 0.842 \\ \hline
0.6 & 1.994 [$\pm 0.033$] & 0.203  \\ \hline
0.7 & 1.973692 [$\pm 0.027$] & 0.266 \\ \hline
0.8 & 1.8706 [$\pm 0.029$] & 0.147\\ \hline
0.9 & 1.8113 [$\pm 0.03$] & 1 \\  \hline
\end{tabular}
\end{center}
\end{table}

\begin{table}
\begin{center}
\caption{PSD of the output curve for a range of Hurst parameters
of the input signal, for Configuration II, $\beta$ = 100, $r\in [5r_g,20r_g]$ \label{tab:C2b100}}
\begin{tabular}{lll}
\hline
H   & $\alpha$            & $p_B$ \\ \hline
\textbf{0.1} & 0.4782 [$\pm 0.030$] & 0.837 \\ \hline
0.2 &  1.1112 [$\pm 0.029$] & 0.455  \\ \hline
0.3 & 1.326146 [$\pm 0.025$] & 0.196 \\ \hline
0.4 & 1.5925 [$\pm 0.031$] & 0.192  \\ \hline
0.5 & 1.7372 [$\pm 0.031$] & 0.41 \\ \hline
0.6 &  1.915 [$\pm 0.03$] & 0.459 \\ \hline
0.7 & 1.8411 [$\pm 0.029$] & 0.995 \\ \hline
\textbf{0.8} & 2.5653 [$\pm 0.031$] & 0.835\\ \hline
0.9 & 1.8059 [$\pm 0.028$] & 0.998 \\  \hline
\end{tabular}
\end{center}
\end{table}

\begin{table}
\begin{center}
\caption{PSD of the output curve for a range of Hurst parameters
of the input signal, for Configuration I, $\beta$ = 1000, $r\in [5r_g,20r_g]$ \label{tab:C1b1000}}
\begin{tabular}{lll}
\hline
H   & $\alpha$            & $p_B$ \\ \hline
0.1 & 0.6786 [$\pm 0.026$] & 0.171  \\ \hline
0.2 & 0.9013 [$\pm 0.03$] & 0.131  \\ \hline
\textbf{0.3} & 1.2751 [$\pm 0.033$] & 0.796 \\ \hline
0.4 & 1.5198 [$\pm 0.031$] & 0.134  \\ \hline
0.5 &  1.7639 [$\pm 0.031$] & 0.402 \\ \hline
0.6 & 1.8806 [$\pm 0.032$] & 1.0 \\ \hline
0.7 & 2.250149 [$\pm 0.028$] & 0.74 \\ \hline
0.8 & 1.8333 [$\pm 0.031$] & 1.0\\ \hline
0.9 & 1.9412 [$\pm 0.026$] & 0.015 \\   \hline
\end{tabular}
\end{center}
\end{table}

\begin{table}
\begin{center}
\caption{PSD of the output curve for a range of Hurst parameters
of the input signal, for Configuration II, $\beta$ = 1000, $r\in [5r_g,20r_g]$ \label{tab:C2b1000}}
\begin{tabular}{lll}
\hline
H   & $\alpha$            & $p_B$ \\ \hline
0.1 & 0.7734 [$\pm 0.029$] & 0.329 \\ \hline
0.2 &  0.8547 [$\pm 0.031$] & 0.083  \\ \hline
0.3 & 1.2811 [$\pm 0.029$] & 0.255 \\ \hline
0.4 & 1.6303 [$\pm 0.029$] & 0.102  \\ \hline
0.5 & 1.771 [$\pm 0.03$] & 0.12 \\ \hline
0.6 & 1.875 [$\pm 0.028$] & 1.0 \\ \hline
0.7 & 2.0087 [$\pm 0.029$] & 0.757\\ \hline
0.8 & 1.8255 [$\pm 0.029$] & 1.0 \\ \hline
0.9 & 2.1951 [$\pm 0.026$] & 0.001\\   \hline
\end{tabular}
\end{center}
\end{table}

\begin{table}
\begin{center}
\caption{PSD of the output curve for a range of Hurst parameters
of the input signal, for Configuration I, $\beta$ = 100, $r\in [5r_g,50r_g]$ \label{tab:C1b100-5-50}}
\begin{tabular}{lll}
\hline
H   & $\alpha$            & $p_B$ \\ \hline
0.1 & 0.699 [$\pm 0.02$] & 0.022  \\ \hline
0.2 & 1 [$\pm 0.02$] & 0.486  \\ \hline
0.3 & 1.332 [$\pm 0.01$] & 0.525 \\ \hline
0.4 & 1.605 [$\pm 0.02$] & 0.09  \\ \hline
0.5 & 1.806 [$\pm 0.02$] & 0.368 \\ \hline
0.6 & 2.016 [$\pm 0.02$] & 0.341  \\ \hline
0.7 & 1.901 [$\pm 0.02$] & 0.506 \\ \hline
0.8 & 1.862 [$\pm 0.02$] & 0.1\\ \hline
\textbf{0.9} & 2.692 [$\pm 0.02$] & 0.774  \\  \hline
\end{tabular}
\end{center}
\end{table}

\begin{table}
\begin{center}
\caption{PSD of the output curve for a range of Hurst parameters
of the input signal, for Configuration I, $\beta$ = 1000, $r\in [5r_g,50r_g]$ \label{tab:C1b1000-5-50}}
\begin{tabular}{lll}
\hline
H   & $\alpha$            & $p_B$ \\ \hline
0.1 & 0.752 [$\pm 0.01$]  & 0.218  \\ \hline
0.2 & 0.936 [$\pm 0.02$]  & 0.311  \\ \hline
\textbf{0.3} & 1.360 [$\pm 0.02$]  & 0.971 \\ \hline
\textbf{0.4} & 1.672 [$\pm 0.02$]  & 0.967  \\ \hline
0.5 & 1.822 [$\pm 0.02$]  & 0.962 \\ \hline
0.6 & 1.928 [$\pm 0.02$]  & 0.794 \\ \hline
0.7 & 1.988 [$\pm 0.01$]  & 0.529 \\ \hline
0.8 & 1.835 [$\pm 0.02$]  & 0.655\\ \hline
0.9 & 1.977 [$\pm 0.01$]  & 0.004 \\   \hline
\end{tabular}
\end{center}
\end{table}

\begin{table}
\begin{center}
\caption{PSD of the output curve for a range of Hurst parameters
of the input signal, for Configuration II, $\beta$ = 100, $r\in [5r_g,50r_g]$ \label{tab:C2b100-5-50}}
\begin{tabular}{lll}
\hline
H   & $\alpha$            & $p_B$ \\ \hline
0.1 & 0.697 [$\pm 0.02$]  & 0.036  \\ \hline
\textbf{0.2} & 1.017 [$\pm 0.02$]  & 0.779  \\ \hline
0.3 & 1.414 [$\pm 0.02$]  & 0.136 \\ \hline
0.4 & 1.551 [$\pm 0.02$]  & 0.227  \\ \hline
0.5 & 1.795 [$\pm 0.02$]  & 0.722 \\ \hline
0.6 & 1.866 [$\pm 0.02$]  & 1 \\ \hline
0.7 & 2.244 [$\pm 0.01$]  & 0.002 \\ \hline
0.8 & 1.812 [$\pm 0.02$]  & 1 \\ \hline
0.9 & 1.798 [$\pm 0.01$]  & 1 \\   \hline
\end{tabular}
\end{center}
\end{table}

\begin{table}
\begin{center}
\caption{PSD of the output curve for a range of Hurst parameters
of the input signal, for Configuration II, $\beta$ = 1000, $r\in [5r_g,50r_g]$ \label{tab:C2b1000-5-50}}
\begin{tabular}{lll}
\hline
H   & $\alpha$            & $p_B$ \\ \hline
0.1 & 0.611 [$\pm 0.02$]  & 0.637  \\ \hline
0.2 & 1.009 [$\pm 0.02$]  & 0.061  \\ \hline
0.3 & 1.404 [$\pm 0.02$]  & 0.477 \\ \hline
0.4 & 1.581 [$\pm 0.02$]  & 0.333   \\ \hline
0.5 & 1.830 [$\pm 0.02$]  & 0.806  \\ \hline
0.6 & 1.853 [$\pm 0.02$]  & 1 \\ \hline
0.7 & 2.255 [$\pm 0.02$]  & 0.665 \\ \hline
0.8 & 1.981 [$\pm 0.02$]  & 0.532 \\ \hline
0.9 & 2.189 [$\pm 0.02$]  & 0.37 \\   \hline
\end{tabular}
\end{center}
\end{table}

\begin{table}
\begin{center}
\caption{PSD of the output curve for a range of Hurst parameters
of the input signal, for Configuration I, $\beta$ = 100, $r\in [5r_g,100r_g]$ \label{tab:C1b100-5-100}}
\begin{tabular}{lll}
\hline
H   & $\alpha$            & $p_B$ \\ \hline
0.1 & 0.720 [$\pm 0.02$] & 0.082  \\ \hline
0.2 & 0.953 [$\pm 0.02$] & 0.015  \\ \hline
0.3 & 1.336 [$\pm 0.01$] & 0.391 \\ \hline
0.4 & 1.567 [$\pm 0.02$] & 0.136  \\ \hline
0.5 & 1.801 [$\pm 0.02$] & 0.039 \\ \hline
0.6 & 2.011 [$\pm 0.02$] & 0.751  \\ \hline
0.7 & 1.873 [$\pm 0.02$] & 1 \\ \hline
0.8 & 1.857 [$\pm 0.02$] & 1\\ \hline
0.9 & 1.810 [$\pm 0.02$] & 1  \\  \hline
\end{tabular}
\end{center}
\end{table}

\begin{table}
\begin{center}
\caption{PSD of the output curve for a range of Hurst parameters
of the input signal, for Configuration I, $\beta$ = 1000, $r\in [5r_g,100r_g]$ \label{tab:C1b1000-5-100}}
\begin{tabular}{lll}
\hline
H   & $\alpha$            & $p_B$ \\ \hline
0.1 & 0.733 [$\pm 0.02$]  & 0.042  \\ \hline
0.2 & 1.089 [$\pm 0.02$]  & 0.465  \\ \hline
0.3 & 1.417 [$\pm 0.02$]  & 0.335 \\ \hline
0.4 & 1.606 [$\pm 0.02$]  & 0.215  \\ \hline
0.5 & 1.851 [$\pm 0.02$]  & 0.036 \\ \hline
0.6 & 1.836 [$\pm 0.02$]  & 1 \\ \hline
0.7 & 2.010 [$\pm 0.02$]  & 0.401 \\ \hline
0.8 & 1.828 [$\pm 0.02$]  & 1\\ \hline
0.9 & 1.828 [$\pm 0.02$]  & 1 \\   \hline
\end{tabular}
\end{center}
\end{table}

\begin{table}
\begin{center}
\caption{PSD of the output curve for a range of Hurst parameters
of the input signal, for Configuration II, $\beta$ = 100, $r\in [5r_g,100r_g]$ \label{tab:C2b100-5-100}}
\begin{tabular}{lll}
\hline
H   & $\alpha$            & $p_B$ \\ \hline
0.1 & 0.685 [$\pm 0.02$]  & 0.096  \\ \hline
\textbf{0.2} & 0.998 [$\pm 0.02$]  & 0.954  \\ \hline
0.3 & 1.365 [$\pm 0.02$]  & 0.55 \\ \hline
0.4 & 1.642 [$\pm 0.02$]  & 0.427  \\ \hline
\textbf{0.5} & 1.746 [$\pm 0.02$]  & 0.754 \\ \hline
0.6 & 2.033 [$\pm 0.02$]  & 0.692 \\ \hline
0.7 & 1.936 [$\pm 0.02$]  & 0.973 \\ \hline
0.8 & 1.907 [$\pm 0.02$]  & 0.629 \\ \hline
0.9 & 1.825 [$\pm 0.02$]  & 1 \\   \hline
\end{tabular}
\end{center}
\end{table}

\begin{table}
\begin{center}
\caption{PSD of the output curve for a range of Hurst parameters
of the input signal, for Configuration II, $\beta$ = 1000, $r\in [5r_g,100r_g]$ \label{tab:C2b1000-5-100}}
\begin{tabular}{lll}
\hline
H   & $\alpha$            & $p_B$ \\ \hline
0.1 & 0.693 [$\pm 0.02$]  & 0.041  \\ \hline
0.2 & 1.067 [$\pm 0.02$]  & 0.042  \\ \hline
0.3 & 1.399 [$\pm 0.02$]  & 0.37 \\ \hline
\textbf{0.4} & 1.531 [$\pm 0.02$]  & 0.977   \\ \hline
0.5 & 1.845 [$\pm 0.02$]  & 0.02  \\ \hline
0.6 & 2.018 [$\pm 0.02$]  & 0.849 \\ \hline
0.7 & 1.843 [$\pm 0.02$]  & 1 \\ \hline
0.8 & 1.827 [$\pm 0.02$]  & 1 \\ \hline
0.9 & 1.883 [$\pm 0.02$]  & 0.11 \\   \hline
\end{tabular}
\end{center}
\end{table}

\end{document}